\documentstyle[12pt]{article}
\begin{document}
\begin{center}{\Large {\bf Axial and off-axial dynamic transitions in 
uniaxially anisotropic Heisenberg ferromagnet: A comparison}}\end{center}

\begin{center}{Muktish Acharyya} \end{center} \begin{center}
{\it Department of Physics, Krishnanagar Government College}\\
{\it P.O.-Krishnanagar, PIN-741101, Dist.- Nadia, West-Bengal, India}\\
{\it E-mail: muktish@vsnl.net}
\end{center}

\vspace {1cm}
\noindent  Uniaxially anisotropic Heisenberg ferromagnet,
in presence of a magnetic field varying sinusoidally in time, is studied
by Monte Carlo simulation. The axial (field applied only along the direction of
anisotropy) and off-axial (field applied only along the direction which is perpendicular
to the direction of anisotropy) dynamic transitions are studied. By studying
the distribution of dynamic order parameter component it is observed that
the axial transition is discontinuous for low anisotropy and becomes continuous in high 
anisotropy. 
The off-axial transition is found to be continuous for all values of anisotropy.
In the infinite anisotropy limit, both types of transitions are compared with that observed
in an Ising ferromagnet for the same value of the field and frequency. 
The infinitely anisotropic axial transition and the dynamic transition in
the Ising ferromagnet occur
at different temperatures whereas the infinitely anisotropic off-axial transition and the
equilibrium ferro-para transition in the Ising model
occur at same temperature.

\vspace {1cm}


\vspace {1cm}

\noindent {\bf {1. Introduction}}

\vskip 0.5cm

\noindent The nonequilibrium dynamical phase transition in magnetic model systems has become an 
interesting field of research in last decades \cite{rmp,rev}. Extensive Monte Carlo simulation
yields some interesting new results in the Ising model \cite{rmp}. These studies \cite{rmp,rev,rik} were able to establish the
significant features of nonequilibrium phase transitions having similarities with well known
equilibrium phase transitions.

However, the Ising model is a special case of general magnetic model, for example, the Heisenberg
model. The Heisenberg model (with ferromagnetic interactions) having uniaxial anisotropy has some
general properties which cannot be found in Ising model. But in the limit of infinite anisotropy,
the Heisenberg model can be mapped into Ising model \cite{mat}. So, the natural expectation is, the Heisenberg
model with uniaxial anisotropy can be studied to have the detailed and general microscopic view
and the results can be checked in the limit of infinite anisotropy (which will give the results
in Ising model). In this case of dynamic transitions, mainly in the magnetic model system in presence
of a magnetic field oscillating sinusoidally in time, the Heisenberg model can serve a better role
than an Ising model. 
It would be quite interesting to know the dynamic response of uniaxially anisotropic Heisenberg
model in presence of a magnetic field applied in different directions.
On the other hand, there is another advantage. The results obtained in the Ising
model is well established \cite{rmp}. These results can be used to check the results obtained in Heisenberg
model by approaching the limit of infinite anisotropy. This prompted to study the dynamic transition in Heisenberg
model with uniaxial and single-site anisotropy. Recently, the dynamic transition was studied \cite{ijmpc} in the uniaxially
anisotropic ferromagnetic Heisenberg model and very interestingly it was observed that the dynamic
symmetry of the order parameter component (along the anisotropy direction) can be broken in presence
of a magnetic field applied along the direction which is perpendicular to the direction of anisotropy. 
This transition was named as {\it off-axial} transition.
The transition 
is found to be continuous and the transition temperature increases as the strength of anisotropy increases. 

So,
the questions naturally arise what would be the difference in the dynamic transitions in presence of
a field applied only along the direction of anisotropy ? How the symmetry breaking takes place ? What
would be the nature (continuous or discontinuous) of the transition ? More interestingly, what
would happen in infinite anisotropic case and in the Ising case ? 
To get the answers of these questions, the dynamic transitions in presence of the axial field 
(i.e., the magnetic field applied only along the direction of anisotropy) and the off-axial field
(i.e., the magnetic field applied only along the direction which is perpendicular to the direction
of anisotropy) are studied
in this paper by Monte Carlo simulation using Metropolis rate. Also, a comparison between axial
and off-axial transitions has been made and the results (in the limit of infinite anisotropy)
for both cases are compared with that observed in
the Ising model.

The uniaxially anisotropic Heisenberg model (with ferromagnetic interaction)
is introduced and explained in section 2. The Monte Carlo simulation technique  
is discussed in the section 3.
The numerical results are reported in the next section and the paper ends with a concluding remarks
in section 5.

\vskip 0.5cm

\noindent {\bf {2. The description of the model}}

\vskip 0.5cm

\noindent The Hamiltonian of a classical anisotropic (uniaxial and single-site) Heisenberg model 
with nearest neighbour ferromagnetic interaction
in the presence of a magnetic field can be written as

\begin{equation}
H = -J \sum_{<ij>} \vec S_i \cdot \vec S_j -D \sum_{i} (S_{iz})^2
-{\vec h} \cdot {\sum_{i} {\vec S_i}},
\end{equation}
\noindent where ${\vec S_i} [S_{ix},S_{iy},S_{iz}]$ represents a classical spin vector of
magnitude unity situated at the $i$-th lattice site. So, $S_{ix}^2+S_{iy}^2+S_{iz}^2 = 1$ is
an equation of a unit sphere.
Classical spin
means, this spin vector can be oriented in any direction in the
vector spin space. $J (> 0)$ is the uniform nearest neighbour 
strength of the ferromagnetic interaction. 
The factor $D$ in the second term is the strength of uniaxial ($z$ here)
anisotropy favouring the spin to be aligned along the $z$-axis. The
last term is the spin-field interaction term, where ${\vec h} [h_x,h_y,h_z]$ is the
externally applied magnetic field (uniform over the space). 
When the magnetic field is applied only along the $\alpha$ - direction,
the magnetic field component $h_{\alpha}$ (may be any one of $x$, $y$ and $z$) is oscillating
sinusoidally
in time and can be written as $h_{\alpha} (t) = h^0_{\alpha} {\rm cos} (\omega t)$,
where $h^0_{\alpha}$ and $\omega$ are the amplitude and angular frequency
($\omega = 2 \pi f$; $f$ is frequency) of the oscillating field
respectively. 
Magnetic field $|\vec h|$ and strength of anisotropy $D$
are measured in the unit of $J$.
The model is defined in a simple cubic lattice of
linear size $L$ with periodic boundary conditions applied in all the three directions.

\vskip 0.5cm

\noindent {\bf {3. The Simulation technique}}

\vskip 0.5cm

\noindent The model, described above, has been studied extensively by Monte Carlo
simulation using the following algorithm \cite{algo}. 
Initial configuration is a  random spin configuration. 
Here, the
algorithm used, can be described as follows.
Two different random numbers $r_1$ and $r_2$ 
(uniformly distributed
between -1 and 1) are chosen in such a way that $R^2=(r_1^2+r_2^2)$ becomes less than or equal to unity.
The set of values of $r_1$ and $r_2$, for which $R^2 > 1$, are rejected.
Now, $u={\sqrt{1-R^2}}$.
Then, $S_{ix}=2ur_1$, $S_{iy}=2ur_2$ and $S_{iz}=1-2R^2$.

Starting from an initial random spin configuration (corresponding to high temperature configuration)
the system is slowly cooled down.  
At any fixed
temperature $T$ (measured in the unit of $J/K_B$) and field amplitude $h^0_{\alpha}$ (measured in the
unit of $J$) a lattice site $i$ has
been chosen randomly (random updating). The value of the spin vector at this randomly chosen site is ${\vec S_i}$ (say). 
The energy of the system is given by the Hamiltonian (equation 1) given above. Now, a test spin vector
${\vec S_i^{\prime}}$ is chosen randomly (described by the algorithm above). For this choice of spin
vector at site $i$ the energy will be
$H^{\prime} = -J \sum_{<ij>} \vec S_i^{\prime} \cdot \vec S_j -D \sum_{i} (S^{\prime}_{iz})^2
-{\vec h} \cdot {\sum_{i} {\vec S_i^{\prime}}}$. The change in energy, associated to this change
in direction of spin vector from ${\vec S_i}$ to ${\vec S^{\prime}_i}$, is $\Delta H = H^{\prime} - H$.
Now, the Monte Carlo method \cite{book,binder} will 
decide how far this change is acceptable. The probability of the change
is given by Metropolis rate \cite{book,binder} (used here)
$W({\vec S_i} \rightarrow {\vec S^{\prime}_i}) = 
{\rm Min} [ 1, {\rm exp}(-\Delta
H/K_BT)]$. This probability will be compared with a random number $R_p$ (say)
between 0 and 1. If $R_p$ does not
exceed $W$, the move (the change ${\vec S_i} \rightarrow {\vec S^{\prime}_i}$) is accepted. 
In this way the spin vector ${\vec S_i}$ is updated.
$L^3$ such random updates of spins, defines one Monte Carlo step per site
(MCSS) and this is considered as the unit of time in this simulation. 
The linear frequency ($f = \omega/{2\pi}$) of the oscillating field
is taken equal to 0.001 and kept constant throughout this simulational study. 
So, 1000 MCSS is required to get
one complete cycle of the oscillating field and consequently
1000 MCSS becomes the time
period ($\tau$) of the applied oscillating magnetic field. 
To calculate any macroscopic quantity, like instantaneous magnetisation
components, the following method was employed.
Starting from an initially random configuration (which corresponds
to a high temperature phase) the system is allowed to be 
stabilised (dynamically)
up to $4 \times 10^4$ MCSS ( i.e., 40 complete cycles of the oscillating 
field) and the averages of various physical quantities are calculated
from further $4 \times 10^4$ MCSS (i.e., averaged over further 40
cycles of the oscillating field). This is quite important to get 
stable hysteresis loop and it is checked that the number of MCSS mentioned above
is sufficient to get stable dynamic phase.
Here the total length of this
simulation for one fixed temperature $T$ is $8 \times 10^4$ MCSS
(which produces 80 complete cycles of the oscillating field). 
Then
the system is slowly 
cooled down (the value of the temperature $T$ has been reduced by 
small interval) to get the values of the statistical quantities in the low
temperature ordered phase. Here, the last spin configuration 
obtained at the previous temperature is used as the initial configuration
for the new temperature. The CPU time required for $8 \times 10^4$ MCSS
is approximately 22 minutes on an Intel-Pentium-III processor.

\vskip 0.5cm

\noindent {\bf {4. Numerical results}}

\vskip 0.5cm

\noindent The linear size of the system $L$ has been taken equal to 20.
The instantaneous magnetisation components (per lattice site)
$m_x ={\sum_{i} S^x_i}/{L^3}, m_y = {\sum_{i} S^y_i}/{L^3}, 
m_z = {\sum_{i} S^z_i}/{L^3}$ are calculated at each time in presence of
magnetic field.
The time averaged (over a full cycle of the oscillating field) 
magnetisation components (the dynamic order parameter components)  
$Q_x = {1 \over \tau} \oint m_x dt, Q_y = {1 \over 
\tau} \oint m_y dt$ and $Q_z = {1 \over \tau} \oint m_z dt$
are calculated by integrating (over the complete cycle of the oscillating field) 
the instantaneous magnetisation components.
The total (vector) dynamic order parameter is expressed as
$\vec Q = {\hat x}Q_x + {\hat y}Q_y + {\hat z}Q_z.$

In this paper, two kinds of dynamic transitions were studied and compared. The {\it axial}
transition means the dynamic order parameter component $Q_z$ becomes zero from
a nonzero value at a finite temperature (the transition temperature) in presence
of a magnetic field $\vec h [0,0,h_z]$ applied only along the direction which is {\it parallel} to the direction
of anisotropy. Since the uniaxial anisotropy has been taken along the z-direction, in this case, 
the direction of magnetic field has only nonzero z-component.
The {\it off-axial} transition \cite{ijmpc} is the transition in presence of
a magnetic field $\vec h [h_x,0,0]$ applied only along the direction which is {\it perpendicular} to the direction of anisotropy. In this case,
the direction of the magnetic field has only nonzero x-component. 

In the case of {\it axial} transition, the instantaneous magnetisation components are
calculated at any fixed temperature $T$, strength of anisotropy $D$ and amplitude of axial
magnetic field $h_z^0$. The time eliminated plot of $m_z-h_z$ gives the axial hysteresis
loop. It was observed that at high temperature ($T = 2.2$) the axial hysteresis loop $m_z-h_z$ is
symmetric (symmetric means the loop is distributed about $h_z$ axis in such a way that the
total z-component of magnetization, over a complete cycle of field, vanishes) (fig.1a). As
a result $Q_z = 0$.  And
at low temperature ($T = 1.0$) the $m_z-h_z$ loop becomes asymmetric ($Q_z \neq 0$) (fig.1b). In both cases,
the $m_x-h_z$ and $m_y-h_z$ loops lie almost along $h_z$ axis, resulting $Q_x$ and $Q_y$
equal to zero respectively. Thus a dynamic transition occurs (as the temperature decreases) at
a certain temperature from a symmetric ($Q_z = 0$; $\vec Q = 0$) to an asymmetric ($Q_z \neq 0$;
$\vec Q \neq 0$) dynamic phase in presence of an {\it axial} magnetic field. 

\vskip 1cm

\begin{center}
\setlength{\unitlength}{0.240900pt}
\ifx\plotpoint\undefined\newsavebox{\plotpoint}\fi
\sbox{\plotpoint}{\rule[-0.200pt]{0.400pt}{0.400pt}}%

\end{center}

\vskip 0.5cm

\noindent {\small {Fig. 1. 
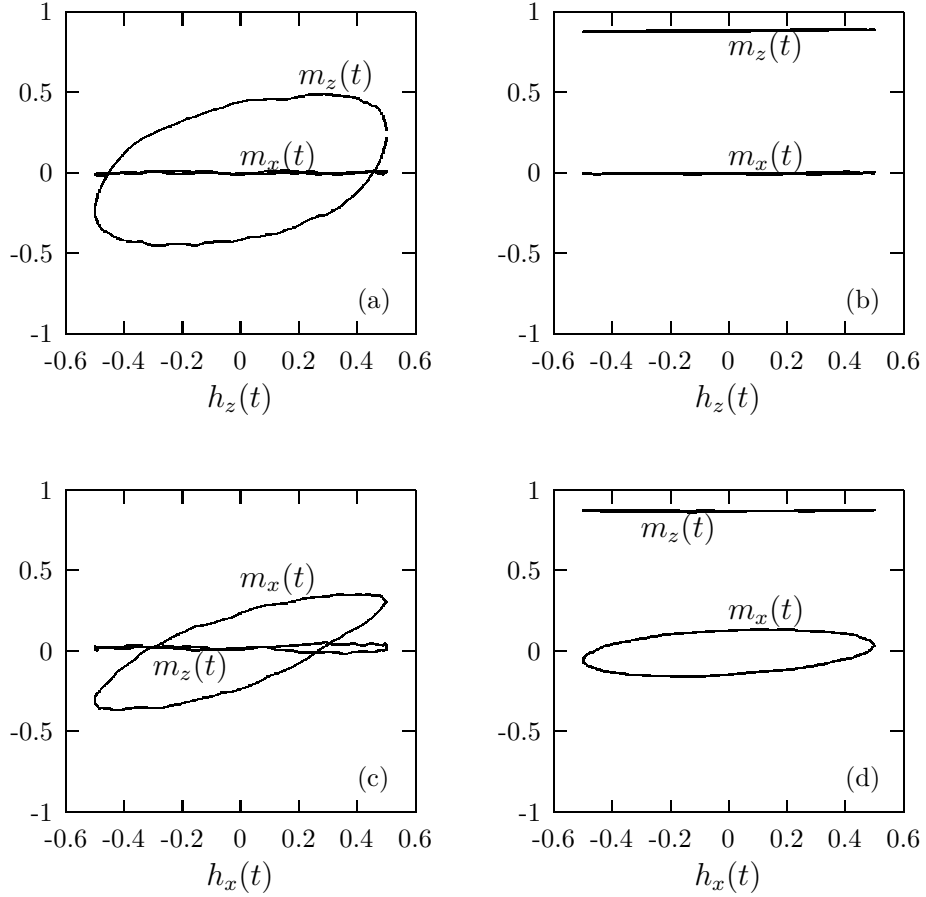
Symmetry breaking in axial and off-axial transitions.
The plot of instantaneous magnetization components against the instantaneous
field components. 
(a) $m_x(t)-h_z(t)$ and $m_z(t)-h_z(t)$ loops for $D = 2.5$, $h_z^0 = 0.5$ and $T = 2.2$,
(b) $m_x(t)-h_z(t)$ and $m_z(t)-h_z(t)$ loops for $D = 2.5$, $h_z^0 = 0.5$ and $T = 1.0$,
(c) $m_x(t)-h_x(t)$ and $m_z(t)-h_x(t)$ loops for $D = 0.5$, $h_x^0 = 0.5$ and $T = 1.8$ and
(d) $m_x(t)-h_x(t)$ and $m_z(t)-h_x(t)$ loops for $D = 0.5$, $h_x^0 = 0.5$ and $T = 0.6$.}}

\vspace {0.5cm}

What was observed in the case of {\it off-axial} transition ? Recently studied \cite{ijmpc} off-axial
transition shows similar dynamic transition via breaking the symmetry of $m_z-h_x$ loop in presence of
an off-axial field (along perpendicular to the anisotropy direction i.e., x-direction). Here, at high
temperature ($T=1.8$) the $m_z-h_x$ loop is symmetric (and $Q_z=0$) and $m_x-h_x$ loop is also
symmetric ($Q_x=0$) (fig.1c). At some lower temperature ($T=0.6$), the $m_z-h_x$ loop becomes asymmetric
($Q_z = 0$) and $m_x-h_x$ loop remains still symmetric ($Q_x = 0$) (fig.1d). In both temperatures
$Q_y = 0$. So, here also a dynamic transition occurs (as the temperature decreases) at a certain temperature
from a symmetric ($Q_z = 0$; $\vec Q = 0$) to an asymmetric ($Q_z \neq 0$; $\vec Q \neq 0$) dynamic phase
in presence of an {\it off-axial} magnetic field. Interestingly, it may be noted here that in higher
temperature the $m_z-h_x$ loop is 'marginally symmetric' (lies very close to $h_x$ axis) 
rather than a symmetric loop (symmetrically distributed away from and about $h_x$ line). Strictly speaking,
the dynamic transition occurs here from a 'marginally symmetric' (loop does not widen up)
to an asymmetric phase. One can differenciate the symmetric phase from the `marginally symmetric' phase
by considering the loop area of that loop whose symmetry breaking is considered in the transition. 
In the symmetric phase loop is sufficiently widen up resulting nonzero
loop area. In Fig. 1a, the $m_z-h_z$ loop area is 0.686 (symmetric loop; $Q_z = 0$). 
But the `marginally symmetric' loops ($m_z-h_x$) have vanishingly small area (0.01)(see Fig. 1c) and 
$Q_z = 0$. It may be noted that,
in the case of off-axial transition, if the magnetic field applied along the x-direction only (oscillating sinusoidally
in time) the $m_x-h_x$ loop is always symmetric (consequently $Q_x = 0$) irrespective of the value of temperature and the strength of
anisotropy $D$ (z-axis). Similarly, for any field applied along y-direction only, the $m_y-h_y$ loop is found to be
always symmetric (i.e., $Q_y = 0$) irrespective of value of $T$ and $D$. But in both cases, whether the off-axial loops
i.e., $m_z-h_x$ or $m_z-h_y$ will be symmetric (rather `marginally symmetric') or asymmetric that
 depends upon the values of temperature $T$, anisotropy $D$
and the magnetic field amplitude $h_x^0$ (or $h_y^0)$. 
These results signify that without anisotropy the dynamic transition (associated to the dynamic symmetry breaking)
cannot be observed in the classical Heisenberg model.
 
\vskip 1 cm

\begin{center}
\setlength{\unitlength}{0.240900pt}
\ifx\plotpoint\undefined\newsavebox{\plotpoint}\fi
\sbox{\plotpoint}{\rule[-0.200pt]{0.400pt}{0.400pt}}%

\end{center}

\vskip 0.5cm

\noindent {\small {Fig. 2. The axial dynamic transitions. 
Temperature ($T$) variations of dynamic order parameter components $Q_z$
for different values of anisotropy strength ($D$) represented by different symbols. $D = 0.5 (\Diamond)$,
$D = 2.5 (+)$, $D = 5.0 (\Box)$, $D = 15.0 (\times)$ and $D = 400.0 (\triangle)$. In all these cases
for the ${\it axial}$ transitions $h_z^0 = 0.5$. 
The data for the temperature variation of dynamic order parameter in 
the Ising model (for $h_z^0 = 0.5$ and $f = 0.001$) are represented by $\star$. Continuous lines in all 
cases are just connecting the 
data points.}}

\vspace {0.6cm}

To investigate the dependence of transition temperature on the strength of anisotropy ($D$) in the case of
axial transition, the temperature variation of dynamic order parameter component $Q_z$ was studied for different
values of $D$. Figure 2 shows the temperature variation of $Q_z$ for different values of $D$. Here, like the
case of off-axial transition \cite{ijmpc} the transition temperature increases as the strength of anisotropy increases.  
It is observed that the axial transition is discontinuous for lower
values of anisotropy strength $D$ (i.e., 0.5, 2.5 etc.) and it becomes continuous for higher values of $D$ (i.e., 5.0 15.0 etc.). 
In the Ising limit ($D \rightarrow \infty$)
the axial transition is also shown in the same figure for $D = 400$). This choice of the value of $D (=400)$ is not arbitrary. In the case
of equilibrium transition it was checked by MC simulation that the value of the magnetisation at any temperature (in the ferromagnetic region)
becomes very close
to that (at that temperature) obtained in the Ising model if the strength of anisotropy is chosen above 300.

\vskip 1cm

\begin{center}
\setlength{\unitlength}{0.240900pt}
\ifx\plotpoint\undefined\newsavebox{\plotpoint}\fi
\sbox{\plotpoint}{\rule[-0.200pt]{0.400pt}{0.400pt}}%
\begin{picture}(900,360)(0,0)
\font\gnuplot=cmr10 at 10pt
\gnuplot
\sbox{\plotpoint}{\rule[-0.200pt]{0.400pt}{0.400pt}}%
\put(121.0,82.0){\rule[-0.200pt]{4.818pt}{0.400pt}}
\put(101,82){\makebox(0,0)[r]{0}}
\put(819.0,82.0){\rule[-0.200pt]{4.818pt}{0.400pt}}
\put(121.0,320.0){\rule[-0.200pt]{4.818pt}{0.400pt}}
\put(101,320){\makebox(0,0)[r]{5}}
\put(819.0,320.0){\rule[-0.200pt]{4.818pt}{0.400pt}}
\put(121.0,82.0){\rule[-0.200pt]{0.400pt}{4.818pt}}
\put(121,41){\makebox(0,0){-1}}
\put(121.0,300.0){\rule[-0.200pt]{0.400pt}{4.818pt}}
\put(301.0,82.0){\rule[-0.200pt]{0.400pt}{4.818pt}}
\put(301,41){\makebox(0,0){-0.5}}
\put(301.0,300.0){\rule[-0.200pt]{0.400pt}{4.818pt}}
\put(480.0,82.0){\rule[-0.200pt]{0.400pt}{4.818pt}}
\put(480,41){\makebox(0,0){0}}
\put(480.0,300.0){\rule[-0.200pt]{0.400pt}{4.818pt}}
\put(660.0,82.0){\rule[-0.200pt]{0.400pt}{4.818pt}}
\put(660,41){\makebox(0,0){0.5}}
\put(660.0,300.0){\rule[-0.200pt]{0.400pt}{4.818pt}}
\put(839.0,82.0){\rule[-0.200pt]{0.400pt}{4.818pt}}
\put(839,41){\makebox(0,0){1}}
\put(839.0,300.0){\rule[-0.200pt]{0.400pt}{4.818pt}}
\put(121.0,82.0){\rule[-0.200pt]{172.966pt}{0.400pt}}
\put(839.0,82.0){\rule[-0.200pt]{0.400pt}{57.334pt}}
\put(121.0,320.0){\rule[-0.200pt]{172.966pt}{0.400pt}}
\put(40,201){\makebox(0,0){$P(Q_z)$}}
\put(301,201){\makebox(0,0)[l]{$T=1.527$}}
\put(767,272){\makebox(0,0)[l]{(a)}}
\put(121.0,82.0){\rule[-0.200pt]{0.400pt}{57.334pt}}
\put(121,82){\usebox{\plotpoint}}
\put(121.0,82.0){\rule[-0.200pt]{17.345pt}{0.400pt}}
\put(193.0,82.0){\rule[-0.200pt]{0.400pt}{23.608pt}}
\put(193.0,180.0){\rule[-0.200pt]{4.336pt}{0.400pt}}
\put(211.0,180.0){\rule[-0.200pt]{0.400pt}{17.586pt}}
\put(211.0,253.0){\rule[-0.200pt]{4.336pt}{0.400pt}}
\put(229.0,153.0){\rule[-0.200pt]{0.400pt}{24.090pt}}
\put(229.0,153.0){\rule[-0.200pt]{4.336pt}{0.400pt}}
\put(247.0,122.0){\rule[-0.200pt]{0.400pt}{7.468pt}}
\put(247.0,122.0){\rule[-0.200pt]{4.336pt}{0.400pt}}
\put(265.0,106.0){\rule[-0.200pt]{0.400pt}{3.854pt}}
\put(265.0,106.0){\rule[-0.200pt]{4.336pt}{0.400pt}}
\put(283.0,106.0){\rule[-0.200pt]{0.400pt}{0.723pt}}
\put(283.0,109.0){\rule[-0.200pt]{4.336pt}{0.400pt}}
\put(301.0,95.0){\rule[-0.200pt]{0.400pt}{3.373pt}}
\put(301.0,95.0){\rule[-0.200pt]{4.095pt}{0.400pt}}
\put(318.0,95.0){\rule[-0.200pt]{0.400pt}{1.204pt}}
\put(318.0,100.0){\rule[-0.200pt]{4.336pt}{0.400pt}}
\put(336.0,91.0){\rule[-0.200pt]{0.400pt}{2.168pt}}
\put(336.0,91.0){\rule[-0.200pt]{4.336pt}{0.400pt}}
\put(354.0,89.0){\rule[-0.200pt]{0.400pt}{0.482pt}}
\put(354.0,89.0){\rule[-0.200pt]{4.336pt}{0.400pt}}
\put(372.0,86.0){\rule[-0.200pt]{0.400pt}{0.723pt}}
\put(372.0,86.0){\rule[-0.200pt]{4.336pt}{0.400pt}}
\put(390.0,83.0){\rule[-0.200pt]{0.400pt}{0.723pt}}
\put(390.0,83.0){\rule[-0.200pt]{4.336pt}{0.400pt}}
\put(408.0,83.0){\rule[-0.200pt]{0.400pt}{0.964pt}}
\put(408.0,87.0){\rule[-0.200pt]{4.336pt}{0.400pt}}
\put(426.0,83.0){\rule[-0.200pt]{0.400pt}{0.964pt}}
\put(426.0,83.0){\rule[-0.200pt]{4.336pt}{0.400pt}}
\put(444.0,83.0){\usebox{\plotpoint}}
\put(444.0,84.0){\rule[-0.200pt]{4.336pt}{0.400pt}}
\put(462.0,83.0){\usebox{\plotpoint}}
\put(462.0,83.0){\rule[-0.200pt]{4.336pt}{0.400pt}}
\put(480.0,83.0){\usebox{\plotpoint}}
\put(480.0,84.0){\rule[-0.200pt]{4.336pt}{0.400pt}}
\put(498.0,84.0){\rule[-0.200pt]{0.400pt}{0.723pt}}
\put(498.0,87.0){\rule[-0.200pt]{4.336pt}{0.400pt}}
\put(516.0,87.0){\rule[-0.200pt]{0.400pt}{1.204pt}}
\put(516.0,92.0){\rule[-0.200pt]{4.336pt}{0.400pt}}
\put(534.0,92.0){\rule[-0.200pt]{0.400pt}{3.854pt}}
\put(534.0,108.0){\rule[-0.200pt]{4.336pt}{0.400pt}}
\put(552.0,89.0){\rule[-0.200pt]{0.400pt}{4.577pt}}
\put(552.0,89.0){\rule[-0.200pt]{4.336pt}{0.400pt}}
\put(570.0,85.0){\rule[-0.200pt]{0.400pt}{0.964pt}}
\put(570.0,85.0){\rule[-0.200pt]{4.336pt}{0.400pt}}
\put(588.0,85.0){\rule[-0.200pt]{0.400pt}{1.445pt}}
\put(588.0,91.0){\rule[-0.200pt]{4.336pt}{0.400pt}}
\put(606.0,91.0){\usebox{\plotpoint}}
\put(606.0,92.0){\rule[-0.200pt]{4.336pt}{0.400pt}}
\put(624.0,92.0){\rule[-0.200pt]{0.400pt}{0.723pt}}
\put(624.0,95.0){\rule[-0.200pt]{4.336pt}{0.400pt}}
\put(642.0,95.0){\usebox{\plotpoint}}
\put(642.0,96.0){\rule[-0.200pt]{4.336pt}{0.400pt}}
\put(660.0,96.0){\rule[-0.200pt]{0.400pt}{2.891pt}}
\put(660.0,108.0){\rule[-0.200pt]{4.095pt}{0.400pt}}
\put(677.0,108.0){\rule[-0.200pt]{0.400pt}{2.650pt}}
\put(677.0,119.0){\rule[-0.200pt]{4.336pt}{0.400pt}}
\put(695.0,118.0){\usebox{\plotpoint}}
\put(695.0,118.0){\rule[-0.200pt]{4.336pt}{0.400pt}}
\put(713.0,118.0){\rule[-0.200pt]{0.400pt}{6.022pt}}
\put(713.0,143.0){\rule[-0.200pt]{4.336pt}{0.400pt}}
\put(731.0,143.0){\rule[-0.200pt]{0.400pt}{15.658pt}}
\put(731.0,208.0){\rule[-0.200pt]{4.336pt}{0.400pt}}
\put(749.0,159.0){\rule[-0.200pt]{0.400pt}{11.804pt}}
\put(749.0,159.0){\rule[-0.200pt]{4.336pt}{0.400pt}}
\put(767.0,82.0){\rule[-0.200pt]{0.400pt}{18.549pt}}
\put(767.0,82.0){\rule[-0.200pt]{13.009pt}{0.400pt}}
\end{picture}

\begin{picture}(900,360)(0,0)
\font\gnuplot=cmr10 at 10pt
\gnuplot
\put(121.0,82.0){\rule[-0.200pt]{4.818pt}{0.400pt}}
\put(101,82){\makebox(0,0)[r]{0}}
\put(819.0,82.0){\rule[-0.200pt]{4.818pt}{0.400pt}}
\put(121.0,320.0){\rule[-0.200pt]{4.818pt}{0.400pt}}
\put(101,320){\makebox(0,0)[r]{2}}
\put(819.0,320.0){\rule[-0.200pt]{4.818pt}{0.400pt}}
\put(121.0,82.0){\rule[-0.200pt]{0.400pt}{4.818pt}}
\put(121,41){\makebox(0,0){-1}}
\put(121.0,300.0){\rule[-0.200pt]{0.400pt}{4.818pt}}
\put(301.0,82.0){\rule[-0.200pt]{0.400pt}{4.818pt}}
\put(301,41){\makebox(0,0){-0.5}}
\put(301.0,300.0){\rule[-0.200pt]{0.400pt}{4.818pt}}
\put(480.0,82.0){\rule[-0.200pt]{0.400pt}{4.818pt}}
\put(480,41){\makebox(0,0){0}}
\put(480.0,300.0){\rule[-0.200pt]{0.400pt}{4.818pt}}
\put(660.0,82.0){\rule[-0.200pt]{0.400pt}{4.818pt}}
\put(660,41){\makebox(0,0){0.5}}
\put(660.0,300.0){\rule[-0.200pt]{0.400pt}{4.818pt}}
\put(839.0,82.0){\rule[-0.200pt]{0.400pt}{4.818pt}}
\put(839,41){\makebox(0,0){1}}
\put(839.0,300.0){\rule[-0.200pt]{0.400pt}{4.818pt}}
\put(121.0,82.0){\rule[-0.200pt]{172.966pt}{0.400pt}}
\put(839.0,82.0){\rule[-0.200pt]{0.400pt}{57.334pt}}
\put(121.0,320.0){\rule[-0.200pt]{172.966pt}{0.400pt}}
\put(40,201){\makebox(0,0){$P(Q_z)$}}
\put(211,261){\makebox(0,0)[l]{$T=1.543$}}
\put(767,284){\makebox(0,0)[l]{(b)}}
\put(121.0,82.0){\rule[-0.200pt]{0.400pt}{57.334pt}}
\put(121,82){\usebox{\plotpoint}}
\put(121.0,82.0){\rule[-0.200pt]{17.345pt}{0.400pt}}
\put(193.0,82.0){\rule[-0.200pt]{0.400pt}{14.213pt}}
\put(193.0,141.0){\rule[-0.200pt]{8.672pt}{0.400pt}}
\put(229.0,141.0){\rule[-0.200pt]{0.400pt}{8.672pt}}
\put(229.0,177.0){\rule[-0.200pt]{8.672pt}{0.400pt}}
\put(265.0,147.0){\rule[-0.200pt]{0.400pt}{7.227pt}}
\put(265.0,147.0){\rule[-0.200pt]{8.672pt}{0.400pt}}
\put(301.0,147.0){\rule[-0.200pt]{0.400pt}{0.723pt}}
\put(301.0,150.0){\rule[-0.200pt]{8.431pt}{0.400pt}}
\put(336.0,127.0){\rule[-0.200pt]{0.400pt}{5.541pt}}
\put(336.0,127.0){\rule[-0.200pt]{8.672pt}{0.400pt}}
\put(372.0,116.0){\rule[-0.200pt]{0.400pt}{2.650pt}}
\put(372.0,116.0){\rule[-0.200pt]{8.672pt}{0.400pt}}
\put(408.0,96.0){\rule[-0.200pt]{0.400pt}{4.818pt}}
\put(408.0,96.0){\rule[-0.200pt]{8.672pt}{0.400pt}}
\put(444.0,96.0){\rule[-0.200pt]{0.400pt}{24.572pt}}
\put(444.0,198.0){\rule[-0.200pt]{8.672pt}{0.400pt}}
\put(480.0,198.0){\rule[-0.200pt]{0.400pt}{9.395pt}}
\put(480.0,237.0){\rule[-0.200pt]{8.672pt}{0.400pt}}
\put(516.0,237.0){\rule[-0.200pt]{0.400pt}{10.118pt}}
\put(516.0,279.0){\rule[-0.200pt]{8.672pt}{0.400pt}}
\put(552.0,127.0){\rule[-0.200pt]{0.400pt}{36.617pt}}
\put(552.0,127.0){\rule[-0.200pt]{8.672pt}{0.400pt}}
\put(588.0,127.0){\rule[-0.200pt]{0.400pt}{0.482pt}}
\put(588.0,129.0){\rule[-0.200pt]{8.672pt}{0.400pt}}
\put(624.0,129.0){\rule[-0.200pt]{0.400pt}{5.059pt}}
\put(624.0,150.0){\rule[-0.200pt]{8.672pt}{0.400pt}}
\put(660.0,150.0){\rule[-0.200pt]{0.400pt}{4.336pt}}
\put(660.0,168.0){\rule[-0.200pt]{8.431pt}{0.400pt}}
\put(695.0,144.0){\rule[-0.200pt]{0.400pt}{5.782pt}}
\put(695.0,144.0){\rule[-0.200pt]{8.672pt}{0.400pt}}
\put(731.0,116.0){\rule[-0.200pt]{0.400pt}{6.745pt}}
\put(731.0,116.0){\rule[-0.200pt]{8.672pt}{0.400pt}}
\put(767.0,82.0){\rule[-0.200pt]{0.400pt}{8.191pt}}
\put(767.0,82.0){\rule[-0.200pt]{8.672pt}{0.400pt}}
\end{picture}

\begin{picture}(900,368)(0,0)
\font\gnuplot=cmr10 at 10pt
\gnuplot
\put(121.0,123.0){\rule[-0.200pt]{4.818pt}{0.400pt}}
\put(101,123){\makebox(0,0)[r]{0}}
\put(819.0,123.0){\rule[-0.200pt]{4.818pt}{0.400pt}}
\put(121.0,368.0){\rule[-0.200pt]{4.818pt}{0.400pt}}
\put(101,368){\makebox(0,0)[r]{8}}
\put(819.0,368.0){\rule[-0.200pt]{4.818pt}{0.400pt}}
\put(121.0,123.0){\rule[-0.200pt]{0.400pt}{4.818pt}}
\put(121,82){\makebox(0,0){-1}}
\put(121.0,348.0){\rule[-0.200pt]{0.400pt}{4.818pt}}
\put(301.0,123.0){\rule[-0.200pt]{0.400pt}{4.818pt}}
\put(301,82){\makebox(0,0){-0.5}}
\put(301.0,348.0){\rule[-0.200pt]{0.400pt}{4.818pt}}
\put(480.0,123.0){\rule[-0.200pt]{0.400pt}{4.818pt}}
\put(480,82){\makebox(0,0){0}}
\put(480.0,348.0){\rule[-0.200pt]{0.400pt}{4.818pt}}
\put(660.0,123.0){\rule[-0.200pt]{0.400pt}{4.818pt}}
\put(660,82){\makebox(0,0){0.5}}
\put(660.0,348.0){\rule[-0.200pt]{0.400pt}{4.818pt}}
\put(839.0,123.0){\rule[-0.200pt]{0.400pt}{4.818pt}}
\put(839,82){\makebox(0,0){1}}
\put(839.0,348.0){\rule[-0.200pt]{0.400pt}{4.818pt}}
\put(121.0,123.0){\rule[-0.200pt]{172.966pt}{0.400pt}}
\put(839.0,123.0){\rule[-0.200pt]{0.400pt}{59.020pt}}
\put(121.0,368.0){\rule[-0.200pt]{172.966pt}{0.400pt}}
\put(40,245){\makebox(0,0){$P(Q_z)$}}
\put(480,21){\makebox(0,0){$Q_z$}}
\put(211,307){\makebox(0,0)[l]{$T=1.600$}}
\put(767,337){\makebox(0,0)[l]{(c)}}
\put(121.0,123.0){\rule[-0.200pt]{0.400pt}{59.020pt}}
\put(121,123){\usebox{\plotpoint}}
\put(121.0,123.0){\rule[-0.200pt]{21.681pt}{0.400pt}}
\put(211.0,123.0){\rule[-0.200pt]{0.400pt}{1.445pt}}
\put(211.0,129.0){\rule[-0.200pt]{4.336pt}{0.400pt}}
\put(229.0,128.0){\usebox{\plotpoint}}
\put(229.0,128.0){\rule[-0.200pt]{4.336pt}{0.400pt}}
\put(247.0,128.0){\rule[-0.200pt]{0.400pt}{0.482pt}}
\put(247.0,130.0){\rule[-0.200pt]{4.336pt}{0.400pt}}
\put(265.0,128.0){\rule[-0.200pt]{0.400pt}{0.482pt}}
\put(265.0,128.0){\rule[-0.200pt]{4.336pt}{0.400pt}}
\put(283.0,128.0){\usebox{\plotpoint}}
\put(283.0,129.0){\rule[-0.200pt]{4.336pt}{0.400pt}}
\put(301.0,128.0){\usebox{\plotpoint}}
\put(301.0,128.0){\rule[-0.200pt]{4.095pt}{0.400pt}}
\put(318.0,127.0){\usebox{\plotpoint}}
\put(318.0,127.0){\rule[-0.200pt]{4.336pt}{0.400pt}}
\put(336.0,127.0){\rule[-0.200pt]{0.400pt}{0.964pt}}
\put(336.0,131.0){\rule[-0.200pt]{4.336pt}{0.400pt}}
\put(354.0,130.0){\usebox{\plotpoint}}
\put(354.0,130.0){\rule[-0.200pt]{4.336pt}{0.400pt}}
\put(372.0,127.0){\rule[-0.200pt]{0.400pt}{0.723pt}}
\put(372.0,127.0){\rule[-0.200pt]{4.336pt}{0.400pt}}
\put(390.0,125.0){\rule[-0.200pt]{0.400pt}{0.482pt}}
\put(390.0,125.0){\rule[-0.200pt]{4.336pt}{0.400pt}}
\put(408.0,123.0){\rule[-0.200pt]{0.400pt}{0.482pt}}
\put(408.0,123.0){\rule[-0.200pt]{4.336pt}{0.400pt}}
\put(426.0,123.0){\usebox{\plotpoint}}
\put(426.0,124.0){\rule[-0.200pt]{4.336pt}{0.400pt}}
\put(444.0,124.0){\rule[-0.200pt]{0.400pt}{4.577pt}}
\put(444.0,143.0){\rule[-0.200pt]{4.336pt}{0.400pt}}
\put(462.0,143.0){\rule[-0.200pt]{0.400pt}{47.698pt}}
\put(462.0,341.0){\rule[-0.200pt]{4.336pt}{0.400pt}}
\put(480.0,333.0){\rule[-0.200pt]{0.400pt}{1.927pt}}
\put(480.0,333.0){\rule[-0.200pt]{4.336pt}{0.400pt}}
\put(498.0,150.0){\rule[-0.200pt]{0.400pt}{44.085pt}}
\put(498.0,150.0){\rule[-0.200pt]{4.336pt}{0.400pt}}
\put(516.0,150.0){\usebox{\plotpoint}}
\put(516.0,151.0){\rule[-0.200pt]{4.336pt}{0.400pt}}
\put(534.0,134.0){\rule[-0.200pt]{0.400pt}{4.095pt}}
\put(534.0,134.0){\rule[-0.200pt]{4.336pt}{0.400pt}}
\put(552.0,127.0){\rule[-0.200pt]{0.400pt}{1.686pt}}
\put(552.0,127.0){\rule[-0.200pt]{8.672pt}{0.400pt}}
\put(588.0,126.0){\usebox{\plotpoint}}
\put(588.0,126.0){\rule[-0.200pt]{4.336pt}{0.400pt}}
\put(606.0,126.0){\usebox{\plotpoint}}
\put(606.0,127.0){\rule[-0.200pt]{4.336pt}{0.400pt}}
\put(624.0,126.0){\usebox{\plotpoint}}
\put(624.0,126.0){\rule[-0.200pt]{4.336pt}{0.400pt}}
\put(642.0,126.0){\rule[-0.200pt]{0.400pt}{0.723pt}}
\put(642.0,129.0){\rule[-0.200pt]{4.336pt}{0.400pt}}
\put(660.0,128.0){\usebox{\plotpoint}}
\put(660.0,128.0){\rule[-0.200pt]{8.431pt}{0.400pt}}
\put(695.0,126.0){\rule[-0.200pt]{0.400pt}{0.482pt}}
\put(695.0,126.0){\rule[-0.200pt]{4.336pt}{0.400pt}}
\put(713.0,125.0){\usebox{\plotpoint}}
\put(713.0,125.0){\rule[-0.200pt]{4.336pt}{0.400pt}}
\put(731.0,124.0){\usebox{\plotpoint}}
\put(731.0,124.0){\rule[-0.200pt]{4.336pt}{0.400pt}}
\put(749.0,123.0){\usebox{\plotpoint}}
\put(749.0,123.0){\rule[-0.200pt]{17.345pt}{0.400pt}}
\end{picture}
\end{center}

\vskip 0.5cm

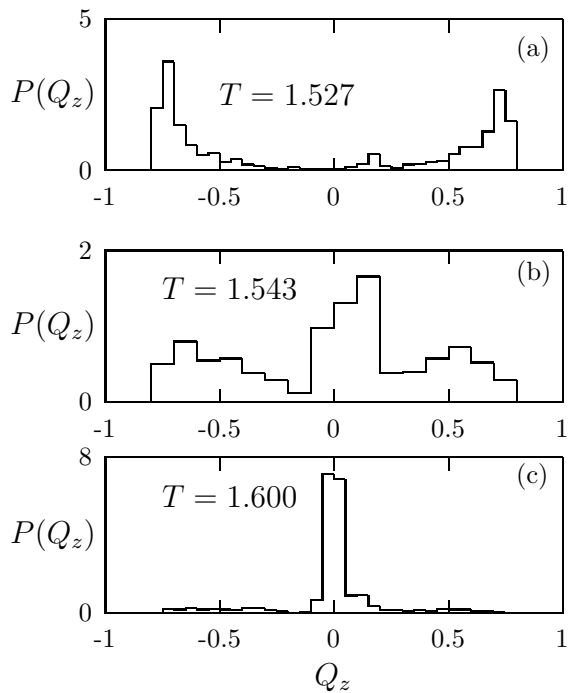
\noindent {\small {Fig. 3. The normalized distributions of dynamic order parameter component $Q_z$ for different
values of temperatures ($T$) in the case of axial transition. Here, $D = 2.5$ and $h_z^0 = 0.5$.}}

\vskip 0.5cm

To establish the discontinuous nature of the transition for lower values of $D$ the well known and widely used method \cite{binder}
i.e., to check the distribution of order
parameter component $Q_z$ at some temperatures very close to transition temperature, was employed here. It may be noted here
that a similar method was
successfully applied recently in the case of dynamic transition in the Ising model \cite{ma}.
Figure 3 shows the normalized distribution ($\int P(Q_z) dQ_z = 1$) of dynamic order parameter component $Q_z$ for different temperatures. 
The distribution ($P(Q_z)$) at each temperature was found from 8000 different values of $Q_z$.
For $D$ = 2.5 the transition occurs around $T \simeq 1.543$ (Fig. 2.).
Below the
transition temperature (i.e., in ordered phase $T = 1.527$) the distribution shows two peaks (fig. 3a) 
and above the transition temperature (i.e., in disordered phase $T = 1.600$) this shows only
one peak (fig.3c). However, very close to the transition temperature ($T \simeq 1.543$) the distribution has three peaks (fig.3b) indicating
clearly the discontinuous nature of the transition. 
The transition becomes continuous for higher values of $D$. The temperature variations of $Q_z$ for infinitely anisotropic ($D = 400$)
 Heisenberg model
and that in the Ising model were compared (for $h_z = 0.5$ and $f = 0.001$ in both cases). In both the cases, the transitions are found to be
continuous. But, the transition temperatures were found to be 
 different (see Fig. 2) although the anisotropic
Heisenberg model maps into an Ising model in $D \rightarrow \infty$ limit.

\vskip 0.75cm

\begin{center}
\setlength{\unitlength}{0.240900pt}
\ifx\plotpoint\undefined\newsavebox{\plotpoint}\fi
\sbox{\plotpoint}{\rule[-0.200pt]{0.400pt}{0.400pt}}%

\end{center}

\vskip 0.5cm

\noindent {\small {Fig. 4. The off-axial dynamic transitions. 
Temperature ($T$) variations of dynamic order parameter components $Q_z$
for different values of anisotropy strength ($D$) represented by different symbols. $D = 0.5 (\Diamond)$,
$D = 2.5 (+)$, $D = 5.0 (\Box)$, $D = 15.0 (\times)$ and $D = 400.0 (\triangle)$. In all these cases
for ${\it off-axial}$ transitions $h_x^0 = 0.5$. 
The data for the zero-field ferro-para
equilibrium Ising transition are represented by $\star$. Continuous lines in all 
cases are just connecting the 
data points.}}

\vskip 0.5cm 


The temperature variations of dynamic order parameter component $Q_z$ in the
case of off-axial transition was also studied and shown in figure 4 for different values of $D$.
This shows that the transition temperature increases as $D$ increases. Here, the
transition is continuous for all values of strength of anisotropy $D$. 
The transition for $D$ = 400 ($D \rightarrow \infty$ limit)
was compared with that in the case of Ising model. This shows that both are continuous and occur at the same point
($T \approx 4.5$) which is very close to the Monte Carlo results of equilibrium ferro-para transition 
temperature ($T_c \simeq 4.511$) \cite{book} in 3-dimensional Ising model.

\vspace{0.75cm}

\noindent {\bf {5. Concluding remarks:}}

The nonequilibrium dynamical phase transition in the uniaxially anisotropic Heisenberg model, 
in presence of magnetic field oscillating sinusoidally in time,
is studied by Monte Carlo simulation.
Two cases were studied in this paper. (i) magnetic field oscillating sinusoidally in time is applied
only along the direction of anisotropy, (ii) magnetic field applied only along the direction perpendicular
to the direction of anisotropy. The transition observed in the first case is named axial and that
corresponding to the second case is called off-axial. A comparative study between axial and off-axial
transition is reported in this paper. Three important aspects are considered here. (a) symmetry
breaking, (b) the order of the transition and (c) the transition in infinite anisotropic limit. 

A dynamic symmetry breaking is observed with this dynamic transition. In the case of axial transition
the dynamic transition occured as the temperature decreases from a symmetric to an asymmetric phase.
Whereas, in off-axial case this symmetry breaking takes place from a 'marginally symmetric' to an asymmetric
phase. The reason behind it is as follows: in the case of axial transition by the application of axial
field (oscillating sinusoidally in time) there is a chance that the spin component along the z-direction
may be reversed in opposit direction which would lead to sufficiently wide and symmetric $m_z-h_z$ loop.
But in the case of off-axial transition it is not possible to reverse 
the z-component of spin by applying 
a field (oscillating sinusoidally in time) perpendicular to the direction of uniaxial anisotropy. In
this case the value of the z-component of magnetisation $m_z$ is almost zero. As a result the 
$m_z-h_x$ loop lies on $h_x = 0$ axis and hence the loop is marginally symmetric.   

In both the cases (axial and off-axial) the transition temperature increases as the strength of anisotropy
increases provided the amplitude of the applied field remains same. The strength of anisotropy tries to align 
the spin vector along the direction of anisotropy. So,
as the strength increases it becomes harder to break the symmetry and consequently more thermal fluctuation
is required to break the symmetry. As a result, the transition temperature increases as the strength of anisotropy
increases. But the difference is the nature of transition. In the case of axial field
the transition is discontinuous for lower values of anisotropy and it becomes continuous for higher values
of anisotropy. The reason behind it is, the axial transition occurs in presence of axial field which reverses
the z-component of magnetisation. So, in lower values of anisotropy the spin vector becomes comparatively more flexible
and the transition occurs mechanically in presence of axial field at lower temperature and it is discontinuous. 
As the anisotropy increases the effect of axial field (of same value) becomes weak and the transition is driven by thermal 
fluctuations and the transition is continuous. In the case of off-axial transition, the off-axial field cannot 
reverse the z-component of magnetisation. But as the value of off-axial field increases, the value of x-component 
of magnetisation increases at the cost of z-component of magnetisation. The transition is driven by thermal 
fluctuations and continuous.  

What will be the situation in the limit of infinite strength of anisotropy ? In the case of axial transition it was observed that
the transition temperature for infinitely anisotropic Heisenberg model differs from that obtained in an Ising model.
Although the equilibrium transitions in infinitely anisotropic Heisenberg model and that in the Ising model gives the same
transition temperature, the nonequilibrium transition temperatures in those two cases are not same. 
Since the magnetic field applied in z-direction oscillating sinusoidally in time, keeps the system always away from
the equilibrium, the system does not become an Ising system even in infinite anisotropy limit. As a result, 
the dynamic transition temperature in infinitely
anisotropic Heisenberg model cannot be same for that obtained in the Ising model.
But in the case of
off-axial transition, the transition temperatures in infinitely anisotropic Heisenberg model and that in the Ising model becomes   
exactly equal. The reason behind it is as follows: in the case of off-axial transition the field is applied perpendicular to the
direction of anisotropy. The effect of axial field oscillating sinusoidally in time has no effect in infinite anisotropic
limit. Though the magnetic field applied in the x-direction oscillating sinusoidally in time, the infinite anisotropic
Heisenberg model becomes an Ising model in statistical and thermal equilibrium.  
Hence, the infinitely anisotropic Heisenberg model in presence of off-axial field maps into the Ising model in zero field.
That is why the nonequilibrium transition in infinitely anisotropic Heisenberg model in presence of off-axial field
and the Ising model (in zero external field) give the same result.

One important point may be noted here regarding the dynamics chosen in this simulation. Since, the spin component
does not commute with the Heisenberg Hamiltonian the spin component has an intrinsic dynamics. Considering this
intrinsic dynamics there was a study \cite{land} about structure factor and transport properties in XY- model.
However, in this paper, the motivation is to study the nonequilibrium phase transition driven by thermal fluctuations. 
To study this, one should choose the dynamics which arises due to the interaction with thermal bath.
Since the objective is different, in this paper, the
dynamics chosen here (which arises solely due to the interaction with thermal bath), is Metropolis dynamics. The 
effect of intrinsic
spin dynamics is not taken into account.

To find the phase boundaries of axial and off-axial transitions, and their dependence on the strength of anisotropy
and the frequency of the oscillating field, is the plan of further study. It is a huge computaional task and requires much
computer time. The work is in progress and the details will be reported elsewhere. 

\vskip 1cm

\noindent {\bf Acknowledgments:} 

\noindent The library facility provided by 
Saha Institute of Nuclear Physics, Calcutta, is gratefully acknowledged.
Author would like to thank the referee for bringing Ref. \cite{algo} into
his notice.

\newpage

\noindent {\large {\bf References:}}

\end{document}